\documentstyle[amssymb,aps,graphicx]{revtex}

\def\ket#1{|#1\rangle}
\def\bra#1{\langle#1|}

\def\be{\begin{equation}}
\def\ee{\end{equation}}

\begin{document}


\title{Estimation of quantum channels with
finite resources}
\author{Markus A. Cirone, Aldo Delgado, Dietmar G. Fischer,
Matthias Freyberger, Holger Mack, and Michael Mussinger}
\address{Abteilung f\"{u}r Quantenphysik,
Universit\"{a}t Ulm, D-89069 Ulm, Germany}

\maketitle

\begin{abstract}
We investigate the problem of determining
the parameters that describe a quantum channel.
It is assumed that the users of the channel have
at best only  partial knowledge of it
and make use of a finite amount
of resources to estimate it.
We discuss simple protocols for the estimation
of the parameters of several classes of channels
that are studied in the current literature.
We define two different quantitative measures
of the quality of the estimation schemes,
one based on the standard deviation, the
other one on the fidelity.
The possibility of protocols that
employ entangled particles is also considered.
It turns out that the use of entangled particles
as a new kind of nonclassical resource
enhances the estimation quality of some classes
of quantum channel. Further, the investigated
methods allow us to extend them to 
higher dimensional quantum systems.
 
\end{abstract}


\section{Introduction}

Quantum information processing has attracted a
lot of interest in recent years, following Deutsch's
investigations \cite{deu} concerning the potentiality
of a quantum computer, i.e., a computer
where information is stored and
processed in quantum systems.
Their application as quantum information carriers gives rise
to outstanding possibilities, like secret communication
(quantum cryptography) and the implementation of
quantum networks and quantum algorithms that are more efficient
than classical ones \cite{bell}.

Many investigations concern the transmission of quantum information
from one party (usually called Alice) to another (Bob)
through a communication channel.
In the most basic configuration the
information is encoded in qubits.
If the qubits are perfectly
protected from environmental influence, Bob receives
them in the same state prepared by Alice.
In the more realistic case, however,
the qubits have a nontrivial dynamics
during the transmission because of their interaction
with the environment \cite{bell}. Therefore,
Bob receives a set of distorted
qubits because of the disturbing
action of the channel.

Up to now investigations have focused mainly on
two subjects: Determination of the channel capacity
\cite{capa} and reconstruction schemes for
the original quantum state under the
assumption that the action of the
quantum channel is known \cite{reco}.
Here we focus our attention on
the problem that precedes,
both from a logical and a practical point of view,
all those
schemes: The problem of determining
the properties of the quantum channel. This problem
has not been investigated so far, with the
exception of very recent articles \cite{fuj2,us}.
The reliable transfer of quantum information requires 
a well known intermediate device.
The knowledge of the behaviour of a 
channel is also essential to construct quantum codes \cite{code,knil}.
In particular, we consider the case
when Alice and Bob use a finite amount $N$
of qubits, as this is
the realistic case.
We assume that Alice and Bob have, if ever,
only a partial knowledge
of the properties of the quantum channel and
they want to estimate the parameters that
characterize it. 

The article is organized as follows. In section \ref{GeneralDescript}
we shall give 
the basic idea of quantum channel estimation and introduce the 
notation as well as the tools to quantify the quality of channel estimation
protocols.  We shall then continue with the problem of parametrizing quantum
channels  appropriately in section \ref{Parametrization}. Then we are in a
position to envisage the  estimation protocol for the case of one parameter
channels in section \ref{OneParameter}. 
In particular, we shall investigate the optimal estimation protocols for the 
depolarizing channel, the phase damping channel and the amplitude damping 
channel.
We shall also give the estimation scheme
for an arbitrary  qubit channel. In section \ref{QubitPauli} we
explore the use
of entanglement as a powerful nonclassical resource in the 
context of quantum channel estimation.
section \ref{QuditPauli} deals with higher dimensional 
quantum channels before we conclude in section \ref{Conclude}.

\section{A General Description of Channel Estimation}
\label{GeneralDescript}
The determination of all properties of a quantum channel is of considerable 
importance for any quantum communication protocol. In practice such a 
quantum channel can be a transmission line, the storage for a quantum 
system, or an uncontrolled time evolution of the underlying quantum 
system. The behaviour of such channels is generally not known from the
beginning, so we have to find methods to gain this knowledge.

This is in an exact way only possible if one has infinite resources, which 
means an infinite amount of well prepared quantum systems. The influence
of the channel on each member of such an ensemble can then be studied,
i.e., the corresponding statistics allows us to characterize the channel.
In a pratical 
application, however, such a condition will never be fulfilled. Instead we have
to come along  with low numbers of available quantum systems. We therefore
cannot  determine the action of a quantum channel perfectly, but only up to
some  accuracy. We therefore speak of channel estimation rather than channel 
determination, which would be the case for infinite resources.

A quantum channel describes the evolution affecting the state of a 
quantum system. It can describe effects like decoherence or interaction 
with the environment as well as  controlled or uncontrolled time evolution 
occuring during storage or transmission. In mathematical terms a quantum 
channel  is a completely positive linear map $\cal C$ (CP-map)
\cite{chan,stine}, which transforms a density operator $\rho$ to
another density operator
\begin{displaymath}
	\rho\,'={\cal C}\rho.
\end{displaymath}
Each quantum channel $\cal C$ can be
parametrized by a vector $\vec\lambda$ with $L$ components. For a
specific channel we shall
therefore write ${\cal C}_{\vec\lambda}$ throughout the paper.  Depending on
the initial knowledge about the channel, the number of  parameters differs. 
The goal of  channel estimation is to specify the 
parameter vector $\vec\lambda$.

\begin{figure}
\begin{center}
\includegraphics[width=8cm]{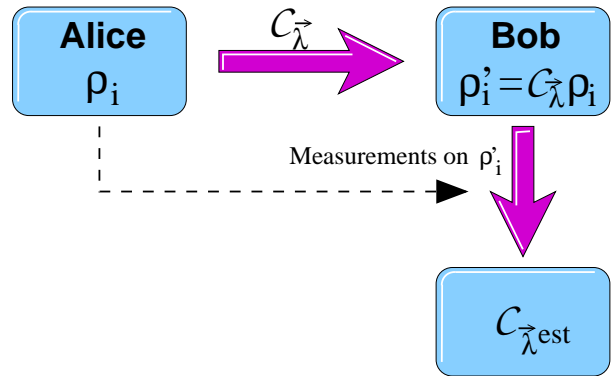}
\caption{Basic scheme of channel estimation. Alice sends $N$ quantum state
$\rho_i$ to Bob. The channel maps these states onto the states $\rho_i'={\cal
C}_{\vec\lambda}\rho_i$, on which Bob can perform arbitrary measurements. Note
that Bob's measurements are designed with the knowledge of the original
quantum states $\rho_i$. His final aim will be to present an estimated vector
$\vec\lambda^{\rm est}$ being as close as possible to the underlying parameter
vector $\vec\lambda$.}
\label{figurescheme}
\end{center}
\end{figure}

The protocol Alice and Bob have to follow in order to estimate the
properties of a quantum channel is depicted in figure \ref{figurescheme}. Alice
and Bob agree on a set of $N$ quantum states $\rho_i, i=1,2,\ldots ,N$, which
are prepared by Alice and then sent through the quantum channel ${\cal
C}_{\vec\lambda}$. Therefore, Bob receives the $N$ states
$\rho\,'_i={\cal C}_{\vec\lambda}\,\rho_i$. He can now perform measurements
on them. From the results he has to deduce an estimated
vector $\vec\lambda^{\rm est}$ which should be as close as possible to the
underlying parameter vector $\vec\lambda$ of the quantum channel.

How can we quantify Bob's estimation? To answer this
we introduce two errors or cost functions which describe how good
the channel is estimated.

The first obvious cost function is the {\em statistical error}
 
\be
  c_s(N,\vec\lambda)\equiv\sum\limits_{\ell=1}^{L}
\left(\lambda_\ell-\lambda_\ell^{\rm est}\left(N\right)\right)^2
\label{ciesse}
\ee
in the estimation of the parameter vector $\vec\lambda$.
Note that the elements of the estimated parameter vector $\vec\lambda^{\rm
est}$ strongly depend on the available resources, i.e.\ the number $N$ of
systems prepared by Alice. We also emphasize that $c_s$ describes the error
for one single run of an estimation protocol. However, we are not interested
in the single run error (\ref{ciesse}) but in the average error of a given
protocol. Therefore, we sum over all possible measurement outcomes $\cal J$ to
get the {\em mean statistical error}
\begin{displaymath}
\overline{c}_s(N,\vec\lambda)\equiv\langle
c_s(N,\vec\lambda)\rangle_{\cal J}
\end{displaymath}
while keeping the number $N$ of resources fixed.

Though this looks as a good benchmark to quantify the quality of
an estimation protocol it has a major drawback. The cost function $c_s$
strongly depends on the parametrization of the quantum channel. While this is
not so important if one compares different protocols using the same
parametrization it anyhow would be much better if we could give a cost
function which is independent of any specifications. We define such a cost
function with the help of the average overlap
\be   {\cal F}({\cal C}_1,{\cal
C}_2)\equiv\Bigl< F({\cal C}_1\,\ket\psi\bra\psi,{\cal
C}_2\,\ket\psi\bra\psi)\Bigr>_{\ket\psi}
\label{channelfidelity}
\ee
between
two quantum channels ${\cal C}_1$ and ${\cal C}_2$, where we average the
fidelity \cite{fid}
\be
F(\rho_1,\rho_2)\equiv{\rm
Tr}^2\,\sqrt{\sqrt{\rho_1} \, \rho_2 \, \sqrt{\rho_1}}
\label{mixedstatefidelity}
\ee
between two mixed states $\rho_1 = {\cal
C}_1\,\rho$ and $\rho_2= {\cal C}_2\,\rho$ over all possible pure quantum states
$\rho = \ket\psi\bra\psi$ \footnote{Since
for our estimation protocol we are only sending pure states through the
quantum channel we also only average over pure states. }.
Since the fidelity ranges from zero to one, the {\em fidelity error}
is given by
\be
c_f(N,\vec\lambda)\equiv1-{\cal F}({\cal
C}_{\vec\lambda},{\cal C}_{\vec\lambda^{\rm est}(N)})
\label{costfct_f}
\ee
which now is zero for identical quantum channels. Again we average over all
possible measurement outcomes to get the  {\em mean fidelity error}
\be  
\overline{c}_f(N,\vec\lambda)\equiv\langle
c_f(N,\vec\lambda)\rangle_{\cal J}
\label{costfct_fav}
\ee
which quantifies the whole protocol and not a specific single run.

In the first part of this paper we are only dealing with qubits
described by the density operator
\be
  \rho\equiv\frac12\left(1+\vec s\cdot\vec\sigma\right)
  \label{blochvector}
\ee
with Bloch vector $\vec s={\rm Tr}(\rho \, \vec\sigma)$ and the Pauli matrices
$\vec\sigma\equiv(\sigma_x,\sigma_y,\sigma_z)$. The action of a channel is
then completely described by the function 
\be   \vec s\,'={\cal C}(\vec s), \label{MapBloch} \ee
which maps the Bloch vector $\vec s$ to a new Bloch vector $\vec s\, '$.
In particular, the mixed state fidelity equation (\ref{mixedstatefidelity}),
for two qubits with Bloch vectors $\vec s_1$ and $\vec s_2$, see
equation (\ref{blochvector}), then simplifies to \cite{fid}
\be
F(\vec s_1,\vec s_2)=\frac12\left[1+\vec s_1\cdot\vec s_2+
\sqrt{\left(1-|\vec s_1|^2\right)\left(1-|\vec s_2|^2\right)}\right]
\label{fidss1}
\ee
which leads to an average channel overlap, equation~(\ref{channelfidelity}),
\be
  {\cal F}({\cal C}_1,{\cal C}_2)\equiv
\int\frac{d\Omega}{4\pi}\ F({\cal C}_1(\vec n),{\cal C}_2(\vec n))
  \label{fidss2}
\ee
for the two qubit channels ${\cal C}_1$ and ${\cal C}_2$.
As emphasized above we only average over pure input states with unit
Bloch vector

\begin{equation}
\vec n=(\cos\Phi\sin\Theta,\sin\Phi\sin\Theta,\cos\Theta)
\label{enne}
\end{equation}
and Bloch sphere element
$d\Omega=\sin\Theta\,d\Theta\,d\Phi$. By inserting equations (\ref{fidss1})
and (\ref{fidss2}) into equations (\ref{costfct_f}) and (\ref{costfct_fav}) we
get a cost function for the comparison of two qubit channels which is
independent of the chosen parametrization.

\section{Parametrization of a quantum channel}
\label{Parametrization}
As we have already mentioned in section \ref{GeneralDescript}, 
the qubits that Alice sends to Bob are fully
characterized by their Bloch vector. Therefore,
the disturbing action of the channel modifies
the initial Bloch vector $\vec{s}$ according to equation (\ref{MapBloch}).
It has been shown that
for completely positive operators the
action of the quantum channel ${\cal C}$ is
given by an affine transformation \cite{fuji}

\begin{equation}
\vec{s}'={\cal M} \; \vec{s}+\vec{v}
\label{eq12}
\end{equation}
where ${\cal M}$ denotes a $3\times 3$ invertible matrix and
$\vec{v}$ is a vector. The transformation is thus
described by 12 parameters, 9 for the matrix ${\cal M}$
and 3 for the vector $\vec{v}$. These 12 parameters
have to fullfill some constraints to guarantee the complete positivity of $\cal
C$ \cite{fuji}.
 The definition
of the parameters is somewhat arbitrary. We will
not use the parameters as defined in \cite{fuji},
but adopt a different parametrization

\begin{equation}
\left(
\begin{array}{c} s'_{1} \\ s'_{2} \\ s'_{3}
\end{array} \right) =
\left(
\begin{array}{ccc}
2\lambda_{7}-\lambda_{1}-\lambda_{4}, \; \;
2\lambda_{10}-\lambda_{1}-\lambda_{4}, \; \;
\lambda_{4}-\lambda_{1} \\
2\lambda_{8}-\lambda_{2}-\lambda_{5}, \; \;
2\lambda_{11}-\lambda_{2}-\lambda_{5}, \; \;
\lambda_{5}-\lambda_{2} \\
2\lambda_{9}-\lambda_{3}-\lambda_{6}, \; \;
2\lambda_{12}-\lambda_{3}-\lambda_{6}, \; \;
\lambda_{6}-\lambda_{3} \\
\end{array} \right)
\left(
\begin{array}{c} s_{1} \\ s_{2} \\ s_{3}
\end{array} \right)
+
\left(
\begin{array}{c}
\lambda_{1}+\lambda_{4}-1 \\
\lambda_{2}+\lambda_{5}-1 \\
\lambda_{3}+\lambda_{6}-1
\end{array} \right)
\label{ssprime}
\end{equation}
in terms of the parameter vector $\vec{\lambda}=
(\lambda_{1},\ldots ,\lambda_{12})^T$.
In general, the choice of the
best parametrization of quantum channels depends on
the relevant features of the channel and also
on which observables can be measured. The reason for the choice
of the parametrization, equation (\ref{ssprime}), will become
clear at the end of the section \ref{OneParameter}, in which a protocol for
the general channel is described.

\section{Estimation of channel parameters}
\label{OneParameter}
Although the characterization of the general quantum channel
requires the determination of 12 parameters, only a smaller
number of parameters must be determined in practice for a
given class of quantum channels. Indeed, the knowledge
of the properties of the physical devices used for quantum
communication gives information on some parameters
and allows to reduce the number of parameters to be estimated.
We shall now examine in detail
some known channels described by only one parameter.

\subsection{The depolarizing channel}

The first channel we consider is the depolarizing channel.
The relation (\ref{eq12}) between the Bloch
vectors $\vec{s}$ and $\vec{s}\,'$ of Alice's and Bob's
qubit, respectively, reduces to the simple form
\begin{displaymath}
\vec{s}_{\lambda}\,'=(1-2\lambda)\vec{s}, 
\end{displaymath}
 where $0 \le \lambda \le 1/2 $
is the only parameter that describes this quantum channel.
This channel is a good model when quantum information is
encoded in the photon polarization that can
change along the transmission fiber
via random rotations  of
the polarization direction.
If Alice prepares the qubit in the pure state  
$\rho= | \psi \rangle \langle \psi |$,
the qubit Bob receives
is described by the state 

\begin{equation}
\rho'=(1-\lambda)
| \psi \rangle \langle \psi |
+\lambda | \bar{\psi} \rangle
\langle \bar{\psi} |,
\label{dep}
\end{equation}
where $\mid \!\! \bar{\psi} \rangle$ denotes
the state orthogonal to
$\mid \!\! \psi \rangle$.
We note that the depolarizing channel has
no preferred basis. Therefore, its action is isotropic
in the direction of the input state.
This means that the action of the channel is
described by equation (\ref{dep}) even after changing
the basis of states.

\begin{figure}
\begin{center}
\includegraphics[width=12cm]{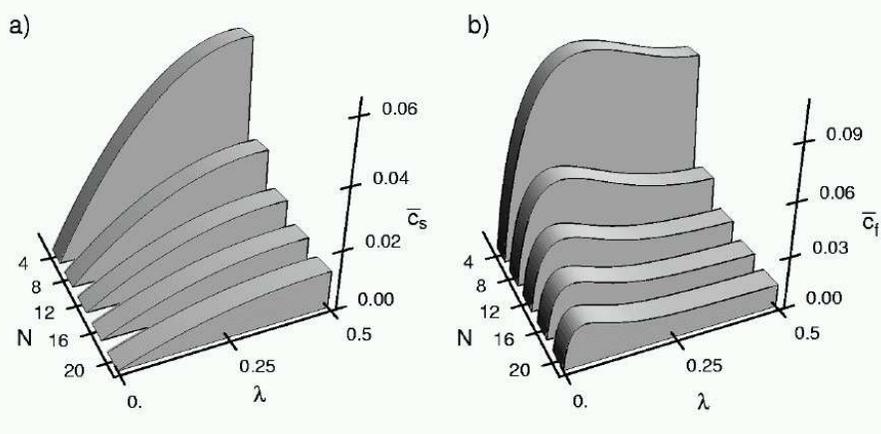}
\caption{a) The average statistical mean error $\bar{c}_s (N, \lambda)$
equation (\ref{eq16}), and b) the fidelity mean error $\bar{c}_f
(N,\lambda)$, equation (\ref{EqDepolarizingCf}), for a depolarizing channel
with parameter $\lambda$, are shown for different  values of the number of
qubits N.}
\label{DepolarizingFigure}
\end{center}
\end{figure}

Bob must estimate $\lambda$, which ranges between 0
(noiseless channel) and 1/2  (total depolarization).
The estimation protocol is the following: Alice
prepares $N$ qubits in the pure state

\begin{displaymath}
\mid \uparrow_{\vec{n}} \rangle
=e^{-i\Phi/2}\cos \frac{\Theta}{2}\mid \uparrow_z \rangle
+e^{i\Phi/2}\sin\frac{\Theta}{2}\mid \downarrow_z \rangle
\end{displaymath}
which denotes spin up in the direction $\vec{n}$,
equation (\ref{enne}), in terms
of eigenvectors $\mid \uparrow_z \rangle$ and
$\mid \downarrow_z \rangle$ of $\sigma_z$.
The state $\mid \uparrow_{\vec{n}} \rangle$
is sent to Bob through the channel. Bob
knows the direction $\vec{n}$ and measures
the spin of the qubit he receives along $\vec{n}$.
The outcome probabilities are

\begin{displaymath}
P(\uparrow_{\vec{n}})=1-\lambda , \;\;\;
P(\downarrow_{\vec{n}})=\lambda.
\end{displaymath}
Since Alice sends a finite number of qubits,
Bob can only determine frequencies of
measurements instead of probabilities.
After Bob has measured $i \le N$ qubits
with spin down and the remaining $N-i$ with spin up,
his estimate of $\lambda$ is
$\lambda^{\rm{est}}=i/N$. Note that in a single
run of the probabilistic estimation method we can
get results $\lambda^{\rm{est}}>1/2$, which is outside the range
of the depolarizing channel. However, for the calculation of the
average errors we do not have to take that into account.
The cost function $c_{s}$, equation (\ref{ciesse}), is
thus

\begin{displaymath}
c_{s}(N,\lambda)=\left( \lambda-\frac{i}{N} \right)^2.
\end{displaymath}
The average error is easily obtained when one considers
that Bob's frequencies of measurements occur according to a 
binomial probability distribution, since
each of the $i$ measurements of spin down occurs
with probability $\lambda$ and each of the spin up
measurements occurs with probability $1-\lambda$.
Therefore, the mean statistical error reads

\begin{eqnarray}
 \bar{c}_{s}(N,\lambda) = \sum_{i=0}^{N}
{ N \choose i}
\lambda^{i}(1-\lambda)^{N-i}
\left( \lambda-\lambda^{\rm{est}}\right)^2 \nonumber \\
= \sum_{i=0}^{N}
{ N \choose i}
\lambda^{i}(1-\lambda)^{N-i}
\left( \lambda-\frac{i}{N} \right)^2=
\frac{\lambda(1-\lambda)}{N}.
\label{eq16}
\end{eqnarray}
This function, shown in figure \ref{DepolarizingFigure} a),
scales with the available finite resources $N$. 
It vanishes when the channel faithfully preserves
 the polarization, $\lambda=0$. The largest average
error occurs for $\lambda=1/2$, when the two actions of preserving and
changing the polarization have the same probability to
occur.

The cost function $c_{f}$ is obtained from equations (\ref{costfct_f})
and (\ref{fidss2})

\begin{displaymath}
c_{f}(N,\lambda) = 1-
\left[ \sqrt{\lambda \frac{i}{N}}+\sqrt{\left(1-\lambda\right)
\left(1-\frac{i}{N}\right)}\right]^2,
\end{displaymath}
which leads to the fidelity mean error

\begin{eqnarray}
\bar{c}_{f}(N,\lambda) & = & \sum_{i=0}^{N}
{ N \choose i}\lambda^{i}
\left( 1-\lambda\right)^{N-i}c_{f}(N,\lambda) \nonumber \\
& = &
1-\frac{1}{N}\sum_{i=0}^{N}
{ N \choose i}\lambda^{i}
\left( 1-\lambda\right)^{N-i}\left( \sqrt{\lambda i}+
\sqrt{(1-\lambda)(N-i)}\right)^2,
\label{EqDepolarizingCf}
\end{eqnarray}
which is shown in figure \ref{DepolarizingFigure} b). Although $\bar{c}_f$
does not show a simple $1/N$--dependency, it clearly decreases for increasing
values of the number of qubits $N$. For a given quality $\bar{c}_{s}$ or
$\bar{c}_{f}$ one can use figure \ref{DepolarizingFigure} and
read off the number $N$ of needed resources.

\subsection{The phase damping channel}

The phase-damping channel acts only on two components of the
Bloch vector, leaving the third one unchanged: 
\begin{displaymath}
\vec{s}_{\lambda}\,\! '=
\left( (1-2\lambda)\; s_{1},(1-2\lambda)\; s_{2},s_{3}
\right)^{T} .
\end{displaymath}
Here $0\le \lambda\le 1/2$ is the damping parameter.
This channel, contrarily to the depolarizing
channel, has a preferred basis. In terms of density matrices,
it transforms the initial state

\begin{equation}
\rho=\left(
\begin{array}{cc}
\rho_{\downarrow \downarrow} \;\; \rho_{\downarrow \uparrow} \\
\rho_{\uparrow \downarrow} \;\; \rho_{\uparrow \uparrow}
\end{array}
\right)
\label{denmat}
\end{equation}
where $\rho_{\downarrow \uparrow}\equiv \langle \downarrow_{z}
\mid \rho \mid \uparrow_{z} \rangle$, etc., into

\begin{displaymath}
\rho'=\left(
\begin{array}{cc}
\rho_{\downarrow \downarrow} & (1-2\lambda)\rho_{\downarrow \uparrow} \\
(1-2\lambda)\rho_{\uparrow \downarrow} & \rho_{\uparrow \uparrow}
\end{array}
\right).
\end{displaymath}
We note that here the parameter $\lambda$ only appears in the off-diagonal 
terms. For this reason, the phase damping channel is
a good model to describe decoherence \cite{phda}. Indeed, a repeated
application of this channel leads to a vanishing
of the off--diagonal terms
in $\rho'$, whereas its diagonal terms are preserved.

\begin{figure}
\begin{center}
\includegraphics[width=12cm]{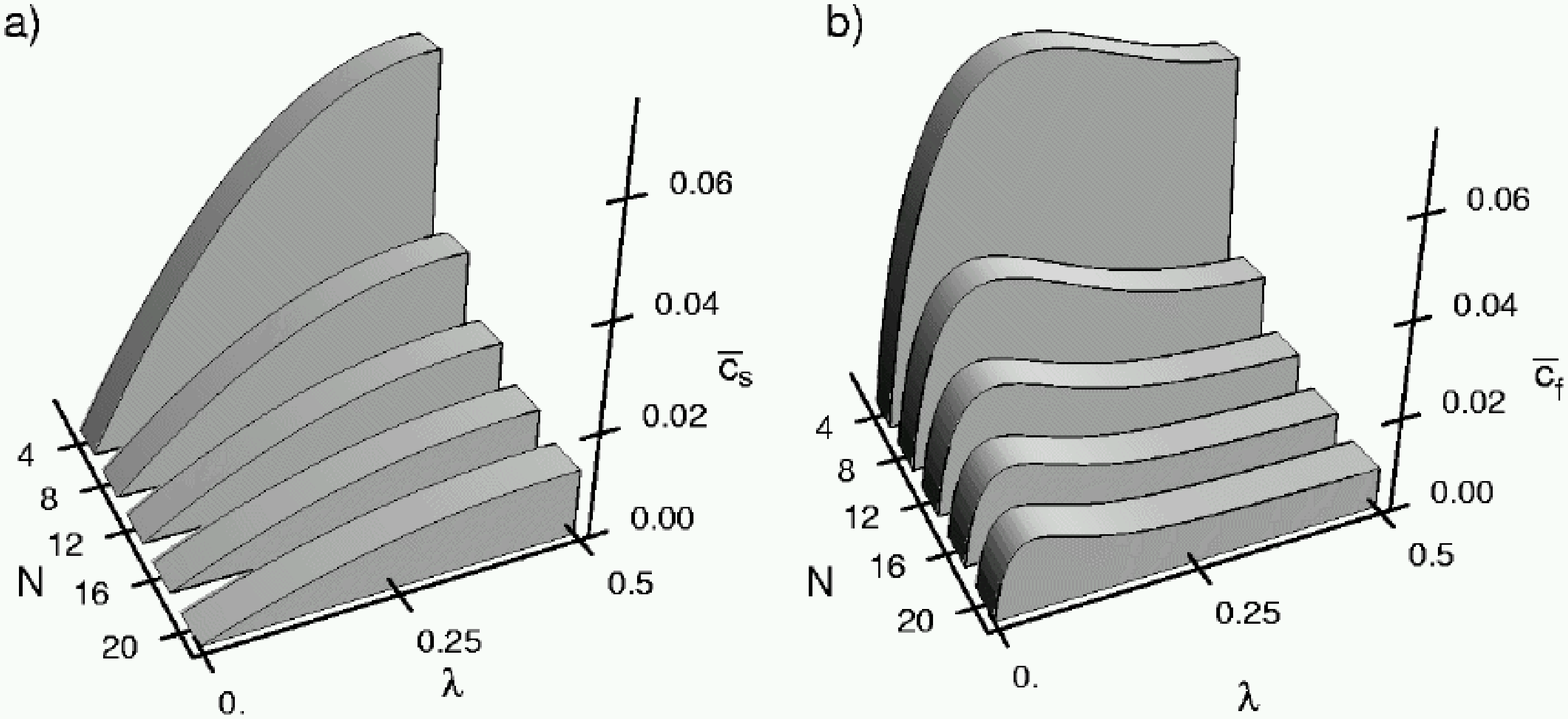}
\caption{a) The average statistical mean error $\bar{c}_s (N, \lambda)$ ,
equation (\ref{eq7}), and b) the fidelity mean error $\bar{c}_f
(N,\lambda)$, equation (\ref{EqPhaseCf}), for a phase damping channel with
parameter $\lambda$, are shown for different  values of the number of qubits
N.}
\label{PhaseDampingFigure}
\end{center}
\end{figure}

Since the parameter $\lambda$ of the
phase damping channel appears in the off-diagonal
elements of $\rho'$, the protocol is the following:
Alice sends $N$ qubits with the Bloch vector
in the $x-y$ plane; for instance, qubits in the state
$\mid \uparrow_{x} \rangle =
(\mid \downarrow_{z} \rangle + \mid \uparrow_{z} \rangle )/\sqrt{2}$
can be used. This would correspond
to a density operator whose
matrix elements are all equal to 1/2, can be used.
The density matrix of Bob's qubit is then given by

\begin{displaymath}
\rho'=\frac{1}{2}\left(
\begin{array}{cc}
1 & (1-2\lambda) \\
(1-2\lambda) & 1
\end{array}
\right).
\end{displaymath}
Now he measures the spin in the $x$ direction.
The theoretical probabilities are

\begin{eqnarray*}
P(\uparrow_{x}) & = &
\langle \uparrow_{x} \mid \rho' \mid \uparrow_{x} \rangle
=1-\lambda, \\
P(\downarrow_{x}) & = &
\langle \downarrow_{x} \mid \rho' \mid \downarrow_{x} \rangle
=\lambda,
\end{eqnarray*}
and we denote the frequency of a spin--down measurement as
$i/N$, leading to $\lambda^{\rm{est}}=i/N$.
The mean statistical error has again the form

\begin{equation}
\bar{c}_{s}(N,\lambda)=\sum_{i=0}^{N}
{ N \choose i}
\lambda^{i}
\left( 1-\lambda \right)^{N-i}
\left( \lambda-\frac{i}{N} \right)^2=\frac{\lambda(1-\lambda)}{N}.
\label{eq7}
\end{equation}
as for the depolarizing channel.
For the fidelity, equation (\ref{fidss1}), we obtain

\begin{eqnarray*}
 F(\vec{s}_{\lambda}\,\! ',\vec{s}\,\, '_{\lambda^{\rm est}}) & = &
1+\left[2 \lambda \lambda^{\rm est} -\lambda -\lambda^{\rm
est} \right. \\
& &\left. +2\sqrt{\lambda(1-\lambda)\lambda^{\rm est}(1-\lambda^{\rm est})}
\right]
\left(s_{1}^{2}+s_{2}^{2}\right)
\end{eqnarray*}
and after averaging over the Bloch surface, according to
equation (\ref{fidss2}),

\begin{displaymath}
{\cal F} ( {\cal C}_{\lambda}, {\cal C}_{\lambda^{\rm est}}) =
1-\frac{2}{3}\lambda-\frac{2}{3}\lambda^{\rm{est}}+
\frac{4}{3}\lambda\lambda^{\rm{est}}+\frac{4}{3}
\sqrt{\lambda(1-\lambda)\lambda^{\rm{est}}
(1-\lambda^{\rm{est}})}
\end{displaymath}
Therefore the fidelity mean error $\bar{c}_{f}$ reads 

\begin{eqnarray}
 \bar{c}_{f}(N,\lambda) = \sum_{i=0}^{N}
{ N \choose i}
\lambda^{i}
\left( 1-\lambda \right)^{N-i}
\frac{2}{3} \left[ \lambda +( 1-2\lambda)\frac{i}{N}
-2\sqrt{\lambda (1-\lambda)\frac{i}{N}\left( 1-\frac{i}{N}\right) } \right]
\nonumber \\
= \frac{4}{3}\left[ \lambda \left( 1-\lambda\right)
-\frac{1}{N}\sum_{i=0}^{N} { N \choose i}
\lambda^{i}\left( 1-\lambda \right)^{N-i} \sqrt{\lambda(1-\lambda)i(N-i)}
\right]
\label{EqPhaseCf}
\end{eqnarray}
which is shown in figure \ref{PhaseDampingFigure} b).
This mean error is very similar to that obtained for the depolarizing channel.

\subsection{The amplitude damping channel}

The amplitude damping channel affects all components of the Bloch
vector according to 

\begin{equation}
\vec{s}_{\lambda}\, '=
\left( \sqrt{1-\lambda}\; s_{1},\sqrt{1-\lambda}\;
s_{2},(1-\lambda)\; s_{3}+\lambda
\right)^{T}
\label{eq28}
\end{equation}
where $0\le \lambda\le 1$ is the damping parameter.
The density matrix, equation (\ref{denmat}), is transformed into

\begin{displaymath}
\rho'=\left(
\begin{array}{cc}
\lambda+(1-\lambda)\rho_{\downarrow \downarrow} &
\sqrt{1-\lambda}\: \rho_{\downarrow \uparrow} \\
\sqrt{1-\lambda}\: \rho_{\uparrow \downarrow} &
(1-\lambda) \rho_{\uparrow \uparrow}
\end{array}
\right) .
\end{displaymath}
This channel is a good model for spontaneous
decay \cite{deca} from an atomic excited state $\mid \uparrow _z\rangle$
to the ground state $\mid \downarrow_z \rangle$.
Repeated applications of this channel cause all elements
but one of the density matrix to vanish.
Now the parameter $\lambda$ appears in all
the elements of $\rho'$ and the channel clearly
possesses a preferred basis.

\begin{figure}
\begin{center}
\includegraphics[width=12cm]{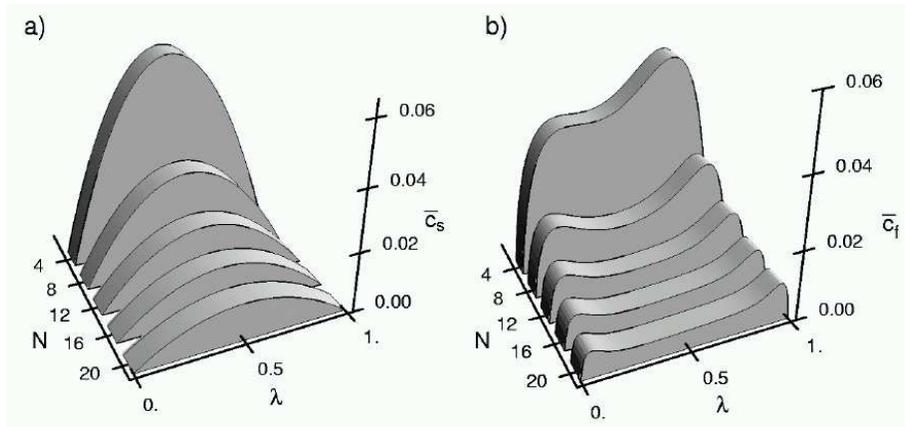}
\caption{a) The average statistical mean error $\bar{c}_s (N, \lambda)$ ,
equation (\ref{eq9}), and b) the fidelity mean error $\bar{c}_f
(N,\lambda)$, equation (\ref{EqAmplitudeCf}), for the amplitude damping channel
with parameter $\lambda$, are shown for different values of the number of
qubits N.}
\label{AmplitudeDampingFigure}
\end{center}
\end{figure}

If Alice and Bob know that they are using an
amplitude damping channel, Alice sends
all $N$ qubits in the state $\mid \uparrow_{z} \rangle$.
The density operator of the qubit received by Bob is 

\begin{displaymath}
\rho'=\left(
\begin{array}{cc}
\lambda & 0 \\
0 & 1-\lambda
\end{array}
\right) .
\end{displaymath}
He measures the spin along the $z$ direction
(the diagonal elements
of $\rho'$ are the probabilities to find
spin down and spin up, respectively). We denote the
frequency of spin down measurements with $i/N$.
The statistical cost function is again

\begin{equation}
\bar{c_{s}}(N,\lambda)=\sum_{i=0}^{N}
{ N \choose i}
\lambda^{i}
(1-\lambda)^{N-i}
\left( \lambda-\frac{i}{N} \right)^2=\frac{\lambda
(1-\lambda)}{N}
\label{eq9}
\end{equation}
Using equation (\ref{eq28}) we obtain

\begin{eqnarray*}
 F(\vec{s}_{\lambda}\, ',\vec{s}\,\,'_{\lambda^{\rm est}}) = \frac{1}{2}
\left[ 1+\sqrt{(1-\lambda)(1-\lambda^{\rm est})}\left(
s_{1}^{2}+s_{2}^{2}\right) +(1-\lambda)(1-\lambda^{\rm est})s_{3}^{2} \right.
\nonumber \\
+
\left[\lambda(1-\lambda^{\rm est})+\lambda^{\rm est}(1-\lambda) \right]s_{3}
\nonumber \\
\left. +\lambda \lambda^{\rm est}+
(1-s_{3})^2\sqrt{\lambda(1-\lambda)}
\sqrt{\lambda^{\rm est}(1-\lambda^{\rm
est})} \right]
\end{eqnarray*}
for the fidelity cost function, equation (\ref{fidss1}), and

\begin{eqnarray*}
{\cal F} ( {\cal C}_{\lambda}, {\cal C}_{\lambda^{{\rm est}}})  & = &
\frac{1}{6}\left[
4+2\sqrt{(1-\lambda)(1-\lambda^{\rm{est}})}-
\lambda-\lambda^{\rm{est}} \right. \nonumber \\
& &\left. +4\lambda\lambda^{\rm{est}}
+4\sqrt{\lambda(1-\lambda)\lambda^{\rm{est}}
(1-\lambda^{\rm{est}})}\right]
\end{eqnarray*}
for the averaged fidelity (\ref{fidss2}). Thus
\begin{eqnarray}
 \bar{c_{f}}(N,\lambda) = \sum_{i=0}^{N}
{ N \choose i}
\lambda^{i}(1-\lambda)^{N-i}
\left[
 1-\frac{1}{6}
\left(
4+2\sqrt{(1-\lambda)\left( 1-\frac{i}{N}\right)}
-\lambda-\frac{i}{N}
\right. \right.
\nonumber \\
\left. \left. +4\lambda\frac{i}{N}
+4\sqrt{\lambda(1-\lambda)\frac{i}{N}
\left( 1-\frac{i}{N}\right)}\:
\right) \right] \nonumber \\
= \frac{1}{3}
\left[
1+\lambda(1-2\lambda)-\sqrt{\frac{1-\lambda}{N}}
\sum_{i=0}^{N}
{ N \choose i } 
\lambda^{N-i}(1-\lambda)^{i}\sqrt{i} \right.
\nonumber \\
\left. -\frac{2}{N}\sqrt{\lambda(1-\lambda)}
\sum_{i=0}^{N}
{ N \choose i}
\lambda^{i}(1-\lambda)^{N-i}\sqrt{i(N-i)} \right]
\label{EqAmplitudeCf}
\end{eqnarray}
which is illustrated by figure \ref{AmplitudeDampingFigure}.

We see from the figures
(\ref{DepolarizingFigure})--(\ref{AmplitudeDampingFigure}) that for a fixed
number of resources the mean error $\bar{c}_{f}$ has a local maximum for small
values of the parameter $\lambda$. For the amplitude damping channel there is
also a second local maximum for large values of $\lambda$.

\subsection{The general quantum channel}

We now come to the problem of estimating the 12 parameters
$\vec{\lambda}=(\lambda_{1},\ldots,\lambda_{12})^T$
of the general quantum channel. The protocol is summarized
in table 1: Alice prepares 12 sets of qubits, divided
into 4 groups. Bob measures the spin of the three sets
in each group along $x$, $y$, and $z$, respectively.
From the measurements on each set he gets an estimate
of one of the parameters $\lambda_{i}$. The parametrization
(\ref{ssprime}) has been chosen for this purpose.

The statistical cost function for the general channel is
thus a generalization of the cost functions for
one--parameter channels,

\begin{eqnarray*}
\bar{c}_{s}(N,\vec{\lambda}) & = & \sum_{i_{1}=0}^{N/12}
\ldots \sum_{i_{12}=0}^{N/12} \left[ \prod_{n=1}^{12}
{ N/12 \choose i_{n} }
\lambda_{n}^{i_{n}}(1-\lambda_{n})^{N/12-i_{n}}\right]
\sum_{j=1}^{12}\left( \lambda_{j}-\lambda_{j}^{\rm{est}} \right)^2
\nonumber \\
& = & \frac{1}{N/12}\sum_{j=1}^{12}\lambda_{j}(1-\lambda_{j}).
\end{eqnarray*}
The mean fidelity error can be calculated numerically.
However, we do not give the expression here
as it gives no particular further insight.

\noindent\parbox{\textwidth}{	
\begin{table}
\caption{The protocol for the general channel. Alice prepares
4 sets of $N/4$ qubits each in a state 
corresponding to the spin pointing in a well prepared direction (left); Bob
splits each set into 3 subsets of $N/12$ qubits each and performs a spin
measurement along $x$, $y$ or $z$, respectively (center); the frequency of
each set of spin measurements gives the estimate of one channel parameter
(right).}
\begin{center}\begin{tabular}{ccc}&
\begin{tabular}{|c|c|c|}
\hline\parbox{25mm}{\centerline{\bf Alice}}&
\parbox{25mm}{\centerline{\bf Bob}}&
\parbox{25mm}{\centerline{\bf Estimate}}\\\hline\hline                
      &$\sigma_z$&$\lambda_{ 1}^{\rm est}$\\\cline{2-3}
$|\downarrow_z\rangle$&$\sigma_x$&$\lambda_{ 2}^{\rm est}$\\\cline{2-3}       
               &$\sigma_y$&$\lambda_{ 3}^{\rm est}$\\\hline\hline             
         &$\sigma_z$&$\lambda_{ 4}^{\rm est}$\\\cline{2-3}
$|\uparrow_z\rangle$  &$\sigma_x$&$\lambda_{ 5}^{\rm est}$\\\cline{2-3}       
               &$\sigma_y$&$\lambda_{ 6}^{\rm est}$\\\hline\hline             
         &$\sigma_z$&$\lambda_{ 7}^{\rm est}$\\\cline{2-3}
$|\uparrow_x\rangle$  &$\sigma_x$&$\lambda_{ 8}^{\rm est}$\\\cline{2-3}       
               &$\sigma_y$&$\lambda_{ 9}^{\rm est}$\\\hline\hline             
         &$\sigma_z$&$\lambda_{10}^{\rm est}$\\\cline{2-3}
$|\uparrow_y\rangle$  &$\sigma_x$&$\lambda_{11}^{\rm est}$\\\cline{2-3}       
               &$\sigma_y$&$\lambda_{12}^{\rm est}$\\\hline
	       \end{tabular}
&\end{tabular}\end{center}
\end{table}
}

\section{The Pauli channel for qubits}
\label{QubitPauli}
Up to now we have only considered estimation methods based
on measuring single qubits sent through the quantum channel.
However, we are not restricted to these estimation schemes.
Instead of sending single qubits we could use entangled
qubit pairs \cite{us}. In this section we will demonstrate
the superiority of such entanglement-based estimation
methods for the estimation of  the so called Pauli
channel by comparing both schemes (for a different use
of entanglement as a powerful resource when using Pauli channels,
see \cite{palma}). \par
The Pauli channel is widely discussed in the literature, especially in the
context of quantum error correction \cite{code}.
Its name originates from the
error operators of the channel. These error operators are the Pauli spin
matrices $\sigma_x$, $\sigma_y$ and $\sigma_z$. The operators define the
quantum mechanical analogue to bit errors in a classical communication channel
since  $\sigma_x$ causes a bit flip
($|\! \downarrow \rangle$ transforms into $| \! \uparrow \rangle$
and vice versa), $\sigma_z$ causes a phase flip 
($| \! \downarrow \rangle+ | \! \uparrow \rangle$ transforms into 
$|\!\downarrow \rangle - |\!\uparrow \rangle $) and $\sigma_y$ results in a
combined bit and phase flip. 
The Pauli channel is described by three probabilities
$\vec{\lambda}\equiv (\lambda_{1},\lambda_{2},
\lambda_{3})^T$ for the occurrence of the three errors. If an initial quantum
state 
$\rho$ is sent through the channel, the Pauli channel transforms $\rho$
into
\begin{equation}
\rho'  =  (1-\lambda_{1}-\lambda_{2}-\lambda_{3})\rho
+\lambda_{1}\sigma_{x}\rho \sigma_{x}
+\lambda_{2}\sigma_{y} \rho \sigma_{y}
+\lambda_{3}
\sigma_{z} \rho \sigma_{z}.
\label{flip}
\label{pau}
\end{equation}
Thus we see that 
the density
operator $\rho$ remains unchanged with probability
$1-\lambda_{1}-\lambda_{2}-\lambda_{3}$, whereas with probability
$\lambda_{1}$ the qubit undergoes a bit flip, 
with probability $\lambda_{3}$ there occurs a phase flip,
and with probability $\lambda_{2}$ both
a bit flip and a phase flip take place.

\subsection{Estimation using single qubits}

We will first describe the single qubit estimation scheme for the Pauli channel,
before we switch to the entanglement-based one in the next subsection.
Following the general estimation scheme presented in
section \ref{GeneralDescript} the protocol to estimate the parameters of the
Pauli channel, equation (\ref{pau}),
requires the preparation of three different quantum states
with spin along three orthogonal directions.
Alice sends (i) $M=N/3$ qubits
in the state $| \uparrow_{z} \rangle$,
(ii) $M$ qubits in the state
$| \uparrow_{x} \rangle$,
and (iii) $M$ qubits in the state
$| \uparrow_{y} \rangle$. Bob measures
their spins along the direction of spin-preparation, namely 
the $z$, $x$, and $y$ axes,
respectively. The measurement probabilities
of spin down along those three
directions are then given by

\begin{eqnarray*}
P(\downarrow_{z}) & = & \lambda_{1}+\lambda_{2}, \\
P(\downarrow_{x}) & = & \lambda_{2}+\lambda_{3}, \\
P(\downarrow_{y}) & = & \lambda_{1}+\lambda_{3}.
\end{eqnarray*}
The estimated parameter values can be calculated via
\begin{eqnarray*}
\lambda_{1}^{\rm{est}} & = &
\frac{1}{2}\left[ \frac{i_{3}}{M}-\frac{i_{1}}{M}
+\frac{i_{2}}{M}\right] \\
\lambda_{2}^{\rm{est}} & = &
\frac{1}{2}\left[ \frac{i_{1}}{M}-\frac{i_{2}}{M}
+\frac{i_{3}}{M}\right] \\
\lambda_{3}^{\rm{est}} & = &
\frac{1}{2}\left[ \frac{i_{2}}{M}-\frac{i_{3}}{M}
+\frac{i_{1}}{M}\right]
\end{eqnarray*}
where $i_{k}/M$, $k=1,2,3$, denote the frequencies of
spin down results along the directions $x$, $y$ and $z$.
We note that, although the probabilities
$\lambda_{i}$ are positive or vanish, their estimated values
$\lambda_{i}^{\rm{est}}$ may be negative. This occurs because
in the present case the measured frequencies are not the
estimates of the parameters.
Nonetheless, the average cost functions 
can always be evaluated.
For the statistical cost function we
find 
\begin{eqnarray}
\bar{c}_s^{\rm \; sep}(N,\vec{\lambda}) & = &
\sum_{i_{1}=0}^{N/3}\sum_{i_{2}=0}^{N/3}\sum_{i_{3}=0}^{N/3}
{ N/3 \choose i_{1} }
{ N/3 \choose i_{2}  }
{ N/3 \choose i_{3} }
(P(\downarrow_{z}))^{i_{3}}
(1-P(\downarrow_{z}))^{N/3-i_{3}}
\nonumber \\
& & \times (P(\downarrow_{x}))^{i_{1}}
(1-P(\downarrow_{x}))^{N/3-i_{1}}
(P(\downarrow_{y}))^{i_{2}}
(1-P(\downarrow_{y}))^{N/3-i_{2}}
\nonumber \\
& & \times \left[
\left( \lambda_{1}-\lambda_{1}^{\rm{est}}
\right)^2+
\left( \lambda_{2}-\lambda_{2}^{\rm{est}}
\right)^2 
+\left( \lambda_{3}-\lambda_{3}^{\rm{est}}
\right)^2
\right] \nonumber \\
& = & \frac{9}{2N}\left[
\lambda_{1}(1-\lambda_{1}-\lambda_{2})+\lambda_{2}(1-\lambda_2-\lambda_3)
\right. \nonumber \\
& & \left. +\lambda_{3}(1-\lambda_3-\lambda_1)\right].
\label{ave}
\end{eqnarray}
for the average error $\bar{c}_s^{\rm \; sep}$ of the estimation with
separable qubits \cite{us}.
For fixed $N$
the average error has a maximum at $\lambda_{1}=\lambda_{2}=
\lambda_{3}=1/4$, in which case all acting operators occur with the same probability. 
On the other hand, the average error vanishes
when faithful transmission or one of the errors
occurs with certainty.\par
Instead of using the statistical cost function $c_s$ we can also use the
fidelity based cost function $c_f$ (\ref{costfct_f}). 
The average error $\bar{c}_f^{\rm \; sep}$ 
of the estimation via separable qubits is then given by
\begin{eqnarray}
\bar{c}_f^{\rm \; sep}(N,\vec{\lambda}) & = & \sum_{i_{1}=0}^{N/3}
\sum_{i_{2}=0}^{N/3} \sum_{i_{3}=0}^{N/3}
{ N/3 \choose i_{1} }
{ N/3 \choose i_{2} }
{ N/3 \choose i_{3} }
(P(\downarrow_{z}))^{i_{3}}
(1-P(\downarrow_{z}))^{N/3-i_{3}}
\nonumber \\
& & \times (P(\downarrow_{x}))^{i_{1}}
(1-P(\downarrow_{x}))^{N/3-i_{1}} \nonumber \\
& & \times (P(\downarrow_{y}))^{i_{2}}
(1-P(\downarrow_{y}))^{N/3-i_{2}}
c_f(N,\vec\lambda).
\label{EqPauliCfsep}
\end{eqnarray}
The cost function $c_f$ cannot be calculated
analytically. We shall compare the results (\ref{ave}) and
(\ref{EqPauliCfsep}) with the same mean errors for
a different estimation scheme, where entangled pairs are used, that we
are now going to illustrate.

\subsection{Estimation using entangled states}

All the protocols that we have considered so far are based
on single qubits prepared in pure states. These qubits are sent
through the channel one after another.
However, one can envisage estimation
schemes with different features.
An interesting and powerful alternative scheme
is based on the use of entangled states \cite{us}. 
In this case the estimation scheme requires
Alice and Bob to share entangled qubits in the
$\mid \! \psi^{-}\rangle\equiv
(\mid \!  \downarrow \rangle \mid \!  \uparrow \rangle
- \mid \!  \uparrow \rangle \mid \!  \downarrow \rangle)/\sqrt{2}$
Bell state. Thus from $N$ initial qubits Alice and Bob can prepare $N/2$ Bell
states. Alice sends her $N/2$ qubits of the entangled pairs
through the Pauli channel, which transforms
the entangled state into the mixed state
\begin{eqnarray*}
\rho'=\cal{C}(| \psi^{-} \rangle \langle \psi^{-} |)
&=& \lambda_{1} | \phi^{-} \rangle
\langle \phi^{-} | +\lambda_{2}
| \phi^{+} \rangle \langle \phi^{+} |
+\lambda_{3} | \psi^{+} \rangle
\langle \psi^{+} | \nonumber \\
& & +\left( 1-\lambda_{1}-\lambda_{2}-\lambda_{3}\right)
| \psi^{-} \rangle \langle \psi^{-} |,
\end{eqnarray*}
where $| \phi^{\pm} \rangle\equiv
\left( | \downarrow \rangle | \downarrow \rangle \pm
| \uparrow \rangle | \uparrow \rangle \right)/\sqrt{2}$
and $| \psi^{\pm} \rangle\equiv
\left( | \downarrow \rangle | \uparrow \rangle \pm
| \uparrow \rangle | \downarrow \rangle \right)/\sqrt{2}$
are the four Bell states \cite{bell}.
Bob performs $N/2$ Bell measurements, i.e., projects each of the $N/2$ 
pairs of qubits 
onto one of the four Bell states with probabilities
\begin{eqnarray*}
P(| \phi^- \rangle )=\lambda_{1}, \;\;\;
P(| \phi^+ \rangle )=\lambda_{2}, \;\;\;
& & P(| \psi^+ \rangle )=\lambda_{3}, \nonumber \\
P(| \psi^- \rangle )=1-\lambda_{1}-\lambda_{2}
-\lambda_{3}.
\end{eqnarray*}
Consequently, the estimated parameter values are directly given by
\begin{displaymath}
\lambda_{1}^{\rm est}=\frac{i_{1}}{N/2},\;\;\;
\lambda_{2}^{\rm est}=\frac{i_{2}}{N/2},\;\;\;
\lambda_{3}^{\rm est}=\frac{i_{3}}{N/2},\;\;\;
\end{displaymath}
We stress that now the estimates of the parameters $\lambda_{i}$ are always
nonnegative, contrarily to the estimation scheme with single
qubits.

The average error $\bar{c}_s^{\rm \; ent}$ of the entangled estimation 
based on the statistical cost function $c_s$  reads
\cite{us}
\begin{eqnarray*}
\bar{c}_s^{\rm \; ent}(N,\vec{\lambda})&=&
\sum_{i_1+i_2+i_3+i_4=N/2}\frac{(N/2)!}{i_1!i_2!i_3!i_4!}\nonumber \\
&\times &
\lambda_1^{i_1}\lambda_2^{i_2}\lambda_3^{i_3}
(1-\lambda_1-\lambda_2-\lambda_3)^{i_4}
\sum_{k=1}^3(\lambda_k-\lambda_k^{est})^2 \nonumber \\
&=& \frac{1}{N/2}\left[\lambda_1(1-\lambda_1)+\lambda_2(1-\lambda_2)
+\lambda_3(1-\lambda_3)\right].
\end{eqnarray*}
As we have shown already in \cite{us} $\bar{c}_s^{\rm \; ent}$ is always
smaller or equal to  $\bar{c}_s^{\rm \; sep}$,

\begin{eqnarray*}
\Delta_{s}(N,\vec{\lambda}) & = &
\bar{c}_s^{\rm \; sep}\left( N, \vec{\lambda} \right)-
\bar{c}_s^{\rm \; ent}\left( N, \vec{\lambda} \right) \nonumber \\
& = & \frac{1}{2N}\left[ 5(1-\lambda_{1}-\lambda_{2}-\lambda_{3})
(\lambda_{1}+\lambda_{2}+\lambda_{3})+\lambda_{1}\lambda_{2} \right.
\nonumber \\
& &\left. +\lambda_{2}\lambda_{3}+\lambda_{3}\lambda_{1}\right]\ge 0
\end{eqnarray*}

\begin{figure}
\begin{center}
\includegraphics[width=8cm]{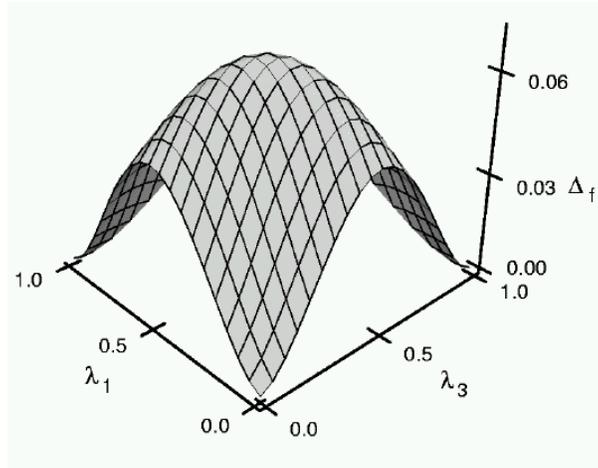}
\caption{The difference $\Delta (N, \vec{\lambda})$ equation (\ref{EqDelta})
between the fidelity mean errors $\bar{c}_f^{\rm sep}$,
equation (\ref{EqPauliCfsep}),  and $\bar{c}_f^{\rm ent}$,
equation (\ref{EqPauliCfent}) for the qubit Pauli channel. $\Delta$ is
plotted   as a function of $\lambda_1$
and $\lambda_3$, while keeping $\lambda_2 = 0$ and $N=6$ fixed.  }
\label{DeltaFigure}
\end{center}
\end{figure}

If we use the fidelity--based cost function $c_f$ we can also write down the
average error $\bar{c}_f^{\rm \; ent}$ for the entangled estimation scheme. It
reads
\begin{eqnarray}
\bar{c}_f^{\rm \; ent}(N,\vec{\lambda})&=&
\sum_{i_1+i_2+i_3+i_4=N/2}\frac{(N/2)!}{i_1!i_2!i_3!i_4!}\nonumber \\
&\times &
\lambda_1^{i_1}\lambda_2^{i_2}\lambda_3^{i_3}
(1-\lambda_1-\lambda_2-\lambda_3)^{i_4}
c_f(N,\vec\lambda)
\label{EqPauliCfent}
\end{eqnarray}
and can be evaluated numerically. 
In figure \ref{DeltaFigure} we show the difference
\begin{equation}
\Delta_f(N,\vec{\lambda})\equiv
\bar{c}_f^{\rm \; sep}\left( N, \vec{\lambda} \right)-
\bar{c}_f^{\rm \; ent}\left( N, \vec{\lambda} \right)
\label{EqDelta}
\end{equation}
between
the average error obtained with single qubits
and entangled qubits. We compare the
two average errors when the same number of
qubits are used. The figure shows that the use
of entangled pairs always leads to an enhanced estimation
and therefore we can consider entanglement as a nonclassical resource
for this application.

Before ending this section we want to add some comments about
our findings. We have found that the mean statistical error has the form

\begin{equation}
\bar{c}_{s}(N,\vec{\lambda})=\frac{L}{N}\sum_{i=1}^{L}\lambda_{i}
(1-\lambda_{i})
\label{univ}
\end{equation}
whenever the estimates of the parameters $\lambda_{i}$ are directly
given by the frequencies of measurements. This occurs also for
the relevant case of the entanglement--based protocol
for the Pauli channel, but not for the qubit--based protocol.
Although the parametrization we have used is better indicated since the
$\lambda_{i}$ represent the probabilities of occurence of the
logic errors, one might be tempted to use a different
parametrization $\vec{\lambda}'$ that gives
again the expression (\ref{univ}). Indeed, this can be done
and actually the errors for the $\vec{\lambda}'$ are smaller.
Nonetheless, one can show that in spite of this improvement, the
scheme based on entangled pairs still gives an enhanced estimation
even with the new parameters $\vec{\lambda}'$. This is not the case
of the one--parameter channels, where the use of entangled pairs
does not lead to enhanced estimation.

\section{The generalized Pauli channel}
\label{QuditPauli}
In the previous section we have proposed an entanglement--based method for
estimating the parameters which define the Pauli
channel. As we shall show in this section, this method can be easily extended
to the case of quantum channels defined on higher dimensional Hilbert spaces.

Let us start by considering the most general possible trace preserving
transformation of a quantum system described initially
by a density operator $\rho$.
This transformation can be written in terms of quantum operations $A_i$ as
\cite{chan}

\begin{equation}
\rho '= {\cal C} \rho = \sum_{i}\lambda_{i}A_{i}\rho A_{i}^{\dagger }
\label{general quantum channel}
\end{equation}

\noindent with  $\sum_{i} \lambda_i A_{i}A_{i}^{\dagger }={\bf 1}$.
 According to the  Stinespring theorem \cite{stine} this is the most
general form of a completely positive linear map. The set of operators
$ A_{i} $ can be interpreted as error operators which
characterize the action of a given quantum channel onto a quantum system
and the set of parameters $ \lambda_{i} $ as
probabilities for the action of error operators $A_{i}$.
 
In particular, we can consider the action of the quantum channel, equation
(\ref {general quantum channel}), onto only one, say the second, of the
subsystems of a bipartite quantum system. For sake of simplicity we assume
that the two subsystems are $D$--level systems. If
only the second particle is affected by the quantum channel,
equation (\ref{general quantum channel}), then the final state is given by 

\begin{equation}
\rho '=\sum_{i}\lambda_{i}\left( {\bf 1\otimes }A_{i}\right) \rho
\left( {\bf 1\otimes }A_{i}\right) ^{\dagger }
\label{quantum channel local}
\end{equation}

\noindent where ${\bf 1}$ is the identity operator.
In the special case of an initially pure state, i. e.
$\rho =\left| \psi \right\rangle \left\langle \psi
\right| $, the final state $\rho '$ becomes

\begin{equation}
\rho '=\sum_{i}\lambda_{i}\left| \psi_i \right\rangle\left\langle\psi_i
\right|   
\label{pure initial state}
\end{equation}

\noindent where we have defined $\left| \psi_i \right\rangle\equiv
\left({\bf 1\otimes }A_{i}\right) \left| \psi \right\rangle $. 

Our aim is the estimation of the parameter values $
\lambda_{i} $ which define the quantum channel. This can be done by
projecting the state of the composite quantum system onto the set of states $%
 \left| \psi_{i}\right\rangle $. As in
the previous protocols, the probabilities $
\lambda_{i} $ can be inferred from the relative frequencies of each
state $%
 \left| \psi_{i} \right\rangle$ and the quality of the estimation can
be quantified with the help of a cost function. However, these states can be 
perfectly distinguished if and only
if they are mutually orthogonal. Hence,
we impose the condition

\begin{equation}
\left\langle \psi_i \right| \left. \psi_j \right\rangle =\left\langle
\psi \right| \left( {\bf 1\otimes }A_{i}\right) ^{\dagger }\left( {\bf %
1\otimes }A_{j}\right) \left| \psi \right\rangle =\delta _{i,j}
\label{condition}
\end{equation}
on the initial state $\left| \psi \right\rangle $.
It means that the initial state $\left| \psi \right\rangle $ of
the bipartite quantum system should be chosen in such a way that it is mapped
onto a set  of mutually orthogonal states $ \left| \psi_i\right\rangle  $ by 
the  error operators $A_{i}$. It is worth to remark
the close analogy between this estimation strategy and quantum error
correction. A non-degenerate quantum error correcting code \cite{knil}
corresponds to a Hilbert subspace which is mapped onto mutually orthogonal
subspaces under the action of the error operators. In this sense, a state
$\left| \psi \right\rangle $ satisfying the condition (\ref{condition}) is a
one-dimensional non-degenerated error correcting code. 

A necessary but not sufficient condition for the existence of a state $%
\left| \psi \right\rangle $ satisfying equation(\ref{condition})
is given by the inequality 

\begin{equation}
n\le D' 
\label{bound}
\end{equation}

\noindent where $n$ is the total 
number of error operators which describe
the quantum channel and $D'$ is the dimension of
the Hilbert space. This
inequality shows why the use of an entangled pair enhances
the estimation scheme. Although the first particle
is not affected by the
action of the noisy quantum channel, it enlarges the dimension
 of the total Hilbert space in such a way that
the condition (\ref
 {condition}) can be fulfilled.
It is also clear from this inequality that if
 the state $\left| \psi
\right\rangle $ is a separable one, a set of orthogonal states can 
 only be designed in principle if the number of error operators $n$ is 
 smaller than $D'$. For instance, in the case of
the Pauli channel for qubits considered in the previous section, condition
(\ref{bound}) does not hold if we use single qubits for the estimation. The
action of the Pauli channel is defined by a set of $n=4$ error operators
(including the identity) acting onto qubits ($D'=2$).

Let us now apply our consideration to the case of a generalized Pauli
channel. The action of this channel is given by

\begin{equation} 
\rho '=\sum_{\alpha,\beta=0}^{D-1}\lambda_{\alpha,\beta} 
U_{\alpha,\beta} \rho \, U_{\alpha,\beta}^{\dagger}.  
\label{generalized pauli channel} 
\end{equation} 

\noindent where the $D^2$
Kraus operators
$U_{\alpha,\beta}$ are defined by

\begin{equation} 
U_{\alpha,\beta}\equiv (X_D)^\beta (Z_D)^\alpha . 
\label{U operators} 
\end{equation}

\noindent The operation

\begin{displaymath}
Z_{D}|j\rangle \equiv e{}^{i\frac{2\pi }{D}j}|j\rangle 
\end{displaymath}
is the generalization of a phase flip in $D$ dimensions. The $D$--dimensional
bit flip is given by
\begin{displaymath}
X_D|j\rangle \equiv |(j-1)~{\rm mod}~D \rangle .
\end{displaymath}
 
\noindent These  errors occur with probabilities $\lambda_{\alpha,\beta}$,
which are normalized,
$\sum_{\alpha,\beta=0}^{D-1}\lambda_{\alpha,\beta}=1$. The states
$|j\rangle, j=0,...,D-1$ form a $D$-dimensional basis of the Hilbert
space of the quantum system. This extension of the Pauli channel to
higher dimensional Hilbert spaces has been studied previously in the context
of quantum error correction \cite{Knill,Gottesmann}, quantum cloning machines
\cite{Cerf} and entanglement \cite{Fivel}. 

Now we will apply the estimation strategy based on condition 
(\ref{condition}) to this particular channel. Thus, we must look for a
state of the bipartite quantum system which satisfies the set of conditions 
(\ref{condition}), for the error operators $U_{\alpha,\beta}$, 
i.e.

\begin{displaymath}
\langle\psi|({\bf 1}\otimes U_{\alpha,\beta})^\dagger
({\bf 1}\otimes U_{\mu,\nu})|\psi\rangle=\delta_{\alpha,\mu}\delta_{\beta,\nu}.
\end{displaymath}

\noindent It can be easily shown that such a state is 

\begin{equation}
|\psi_{0,0}\rangle\equiv \frac{1}{\sqrt{D}}
\sum_{i=0}^{D-1}|i\rangle_1\otimes|i\rangle_2
\label{state for the estimation}
\end{equation}

\noindent where $|i\rangle_1$  and
$|i\rangle_2$ , $j=0,...,D-1$, are basis states for the two particles.
In fact, the
action of the operators ${\bf 1}\otimes U_{\alpha,\beta}$ onto  state
$|\psi_{0,0}\rangle$ generates the basis  

\begin{displaymath}
|\psi_{\alpha,\beta}\rangle\equiv 
{\bf 1}\otimes U_{\alpha,\beta}|\psi_{0,0}\rangle=
\frac{1}{\sqrt{D}}\sum_{k=0}^{D-1}e^{i\frac{2\pi}{D}\alpha k}
|k\rangle_1\otimes|k-\beta\rangle_2
\end{displaymath}

\noindent
of $D^2$ maximally entangled states for the
total Hilbert space.
Thereby, the action of the generalized Pauli channel yields

\begin{displaymath}
\rho '=\sum_{\alpha,\beta=0}^{D-1}\lambda_{\alpha,\beta}
|\psi_{\alpha,\beta}\rangle\langle\psi_{\alpha,\beta}|
\end{displaymath}

\noindent when the initial state is 
$\rho=|\psi_{0,0}\rangle\langle\psi_{0,0}|$. The particular values of the
coefficients $\lambda_{\alpha,\beta}$ can now be obtained by projecting onto
the states  $|\Psi_{\alpha,\beta}\rangle$. The quality of this estimation
strategy according to the statistical error is given by

\begin{displaymath}
\bar{c}_s(N,\vec{\lambda})=\frac{1}{N/2}\sum_{\alpha,\beta=0}^{D-1}
\lambda_{\alpha,\beta}(1-\lambda_{\alpha,\beta})
(1-\delta_{\alpha,0}\delta_{\beta,0})
\end{displaymath}

\noindent with $N/2$ the number of pairs of quantum systems used in the
estimation.

\section{Conclusions}
\label{Conclude}
We have examined the problem of determining the parameters that
describe a noisy quantum channel with finite resources. We have
given simple protocols for the determination of the parameters of several
classes of quantum channels. These protocols are based on
measurements made on qubits that are sent through the channel.
We have also introduced two cost functions that
estimate the quality of the protocols.
In the most simple protocols measurements are performed on each qubit.
We have also shown that more complex schemes based on entangled pairs
can give a better estimate of the parameters of the Pauli channel.
Our investigations stress once more the usefulness of entanglement
in quantum information.

\section*{Acknowledgments}

We acknowledge support by the DFG programme
"Quanten-Informationsverarbeitung", by the European
Science Foundation QIT programme and by the programmes
"QUBITS" and "QUEST" of the European Commission. M~A~C~wishes to thank
V~Bu\v{z}ek for interesting discussions and remarks.


\end{document}